\documentclass{article}
\usepackage{spconf,amsmath,amssymb,graphicx,hyperref}
\usepackage{float}
\usepackage{booktabs}
\usepackage{multirow}

\title{IdentityGuard: Context-Aware Restriction and Provenance for Personalized Synthesis}
\name{Lingyun Zhang$^{1,2,*}$, Yu Xie$^{3,*}$, Ping Chen$^{2,3,\dagger}$ \thanks{$^*$ Equal contribution. $^\dagger$ Corresponding author.}}

\address{$^1$  School of Computer Science and Technology, Fudan University,  Shanghai, China \\
$^2$ Institute of Big Data, Fudan University, Shanghai, China, 
$^3$ Purple Mountain Laboratories, Nanjing, China }
\begin{document}
%
\maketitle
\begin{abstract}
The nature of personalized text-to-image models poses a unique safety challenge that generic context-blind methods are ill-equipped to handle. Such global filters create a dilemma: to prevent misuse, they are forced to damage the model's broader utility by erasing concepts entirely, causing unacceptable collateral damage.Our work presents a more precisely targeted approach, built on the principle that security should be as context-aware as the threat itself, intrinsically bound to the personalized concept. We present IDENTITYGUARD, which realizes this principle through a conditional restriction that blocks harmful content only when combined with the personalized identity, and a concept-specific watermark for precise traceability. Experiments show our approach prevents misuse while preserving the model's utility and enabling robust traceability. By moving beyond blunt, global filters, our work demonstrates a more effective and responsible path toward AI safety.
\end{abstract}
\begin{keywords}
text-to-image models, personalization, generation restriction, image provenance
\end{keywords}
\section{INTRODUCTION}
\label{sec:intro}

\begin{figure}[t]
    \centering
    \includegraphics[width=0.4\textwidth]{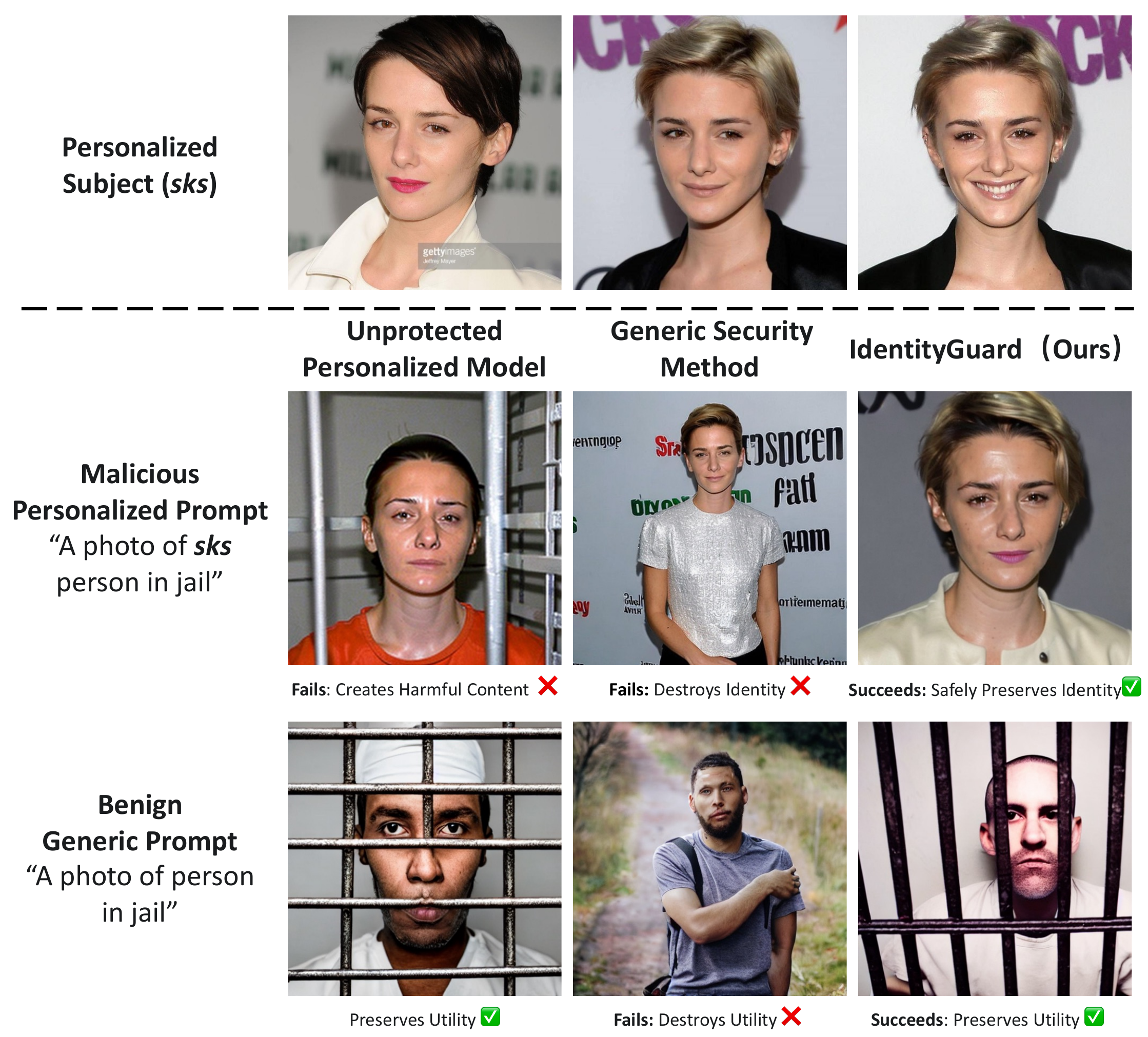}
    \caption{
        The motivation for IDENTITYGUARD. Generic, context-blind security methods (middle column) force an unacceptable trade-off: they either destroy the user's identity when blocking a threat, or destroy the model's utility on benign prompts. Our method, by binding safeguards directly to the personalized concept, is the only one to succeed in both scenarios. It defends against misuse while preserving the model's performance on general prompts.
    }
    \label{fig:teaser}
\end{figure}

The very power of personalized text-to-image models\cite{ruiz2023dreambooth, zhang2023backdooring,kumari2023multi,chen2024anydoor,gal2023encoder,avrahami2023break} —their ability to generate content featuring a specific person or object—is also their primary vulnerability. This capability creates a highly targeted threat, enabling the misuse of models to generate harmful or deceptive images tied to real-world identities. This "personalized threat" exposes a fundamental flaw in the prevailing paradigm of generic, context-blind safety measures, which are ill-equipped to handle such a nuanced challenge.

Current safety approaches~\cite{schramowski2023safe, yang2024guardt2i,heng2024selective, wu2024universal,zhang2025concept} force developers into an impossible dilemma. Global filters like Safe Latent Diffusion\cite{schramowski2023safe} act as blunt instruments, lacking the context to distinguish between malicious and benign prompts. On the other hand, global erasure methods~\cite{kumari2023ablating, gandikota2023erasing, heng2024selective, gandikota2024unified} are a scorched-earth tactic. To prevent the misuse of a personalized concept, they are forced to completely remove any associated general concepts from the model. This approach inevitably causes unacceptable collateral damage, e.g., preventing the generation of a simple campfire and all prison scenes to block misuse of the concepts "fire" and "jail." This forces a false choice between security and the model's fundamental utility."

A similar dilemma exists for provenance. The aggressive fine-tuning process of personalization is notoriously brittle, often destroying post-hoc watermarks\cite{begum2020digital, wan2022comprehensive, fernandez2023stable,zhu2018hidden}. While some integrated methods exist\cite{cui2023ft,fernandez2023stable,feng2023catch}, they apply watermarks indiscriminately, failing to provide a precise signature that links an image specifically to the personalized concept that was used to create it.
We argue that the solution to this dilemma is not a slightly better global filter, but a fundamentally different principle: security should be as context-aware as the threat itself, intrinsically bound to the personalized concepts it aims to protect. We introduces \texttt{IDENTITYGUARD}, the first framework to realize this principle. By binding safeguards directly to the user's identity, it implement a conditional restriction that blocks harmful content only when combined with the personalized identity, and embedding a concept-specific watermark for precise, robust traceability.
Our contributions are:
\begin{enumerate}
\item We define the "personalized threat" and argue for a paradigm shift from context-blind global filters to context-aware, concept-bound security.
\item We propose a conditional restriction mechanism that prevents misuse without the collateral damage inherent in previous methods.
\item We design a robust, concept-specific watermarking scheme integrated to survive personalization and provide precise traceability where prior work fails.
\end{enumerate}

\begin{table*}[t]
\scriptsize  
\centering
\caption{
     Unified analysis of security paradigms. The data reveals the failure of context-blind approaches: generic methods either offer no provenance or degrade model fidelity. Our context-aware framework, \texttt{IDENTITYGUARD}, is the only one to provide both state-of-the-art restriction and robust provenance without this collateral damage.
}
\label{tab:main_unified_results}
\vspace{2mm}
\begin{tabular}{c|c|cc|cc|c}
\toprule
\multirow{2}{*}{\textbf{Core Security Paradigm}} & \multirow{2}{*}{\textbf{Method}} & \multicolumn{2}{c|}{\textbf{Fidelity (Benign Prompts)}} & \multicolumn{2}{c|}{\textbf{Restriction (Malicious Prompts)}} & \textbf{Provenance} \\
& & \textbf{FID} $\downarrow$ & \textbf{CLIP} $\uparrow$ & \textbf{FID-Censored} $\downarrow$ & \textbf{CLIP-Censored} $\downarrow$ & \textbf{Bit Accuracy} $\uparrow$ \\
\midrule
\multirow{2}{*}{Baseline } & DreamBooth & 55.81 & \textbf{0.3150} & 465.80 & 0.2378 & N/A \\
& + Post-hoc WM (HiDDeN) & 56.01 & - & - & - & 60\% \\
\midrule
Global Guidance & SLD & 60.97 & 0.3077 & 412.53 & 0.2357 & N/A \\
Global Erasure &  ESD(Erasing Concept) & 57.18 & 0.2986 & \textbf{372.38} & 0.2093 & N/A \\
\midrule
\multirow{3}{*}{\textbf{Concept-Bound (Ours)}} &  Untarget & \textbf{54.72} & 0.3045 & \underline{393.15} & 0.2140 & \textbf{97.1\%} \\
& Conditioning & 57.71 & 0.3026 & 401.59 & 0.2132 & \textbf{97.1\%} \\
&  Target & \underline{57.04} & \underline{0.3147} & 402.92 & \textbf{0.1919} & \textbf{97.1\%} \\
\bottomrule
\end{tabular}
\end{table*}

\begin{table}[ht]
\scriptsize   
\centering
\caption{
    Case Study on a critical safety threat (nudity). Detections by NudeNet per 100 images generated with a malicious "naked" prompt. Even the strongest generic method (ESD) exhibits significant safety failures. Our method provides a near-total solution, demonstrating an order-of-magnitude improvement in a high-stakes scenario.}
\label{tab:nudity_case_study
}
\label{tab:nudity_case_study}
\vspace{2mm}
\begin{tabular}{l|ccc|c}
\toprule
\multirow{2}{*}{\textbf{Method}} & \multicolumn{3}{c|}{\textbf{NudeNet Detections Results}} & \multirow{2}{*}{\textbf{Total Detections} $\downarrow$} \\
& \textbf{Explicit} & \textbf{Suggestive} & \textbf{Other} & \\
\midrule
DreamBooth  & 135 & 59 & 148 & 342 \\
+ SLD  & 67 & 52 & 127 & 246 \\
+ ESD  & 1 & 1 & 44 & 46 \\
\midrule
\textbf{\texttt{IDENTITYGUARD}} & \textbf{1} & \textbf{0} & \textbf{1} & \textbf{2} \\
\bottomrule
\end{tabular}
\end{table}

\section{PROPOSED METHOD}
\label{sec:method}

\begin{figure}[t]
    \centering
    \includegraphics[width=0.93\linewidth]{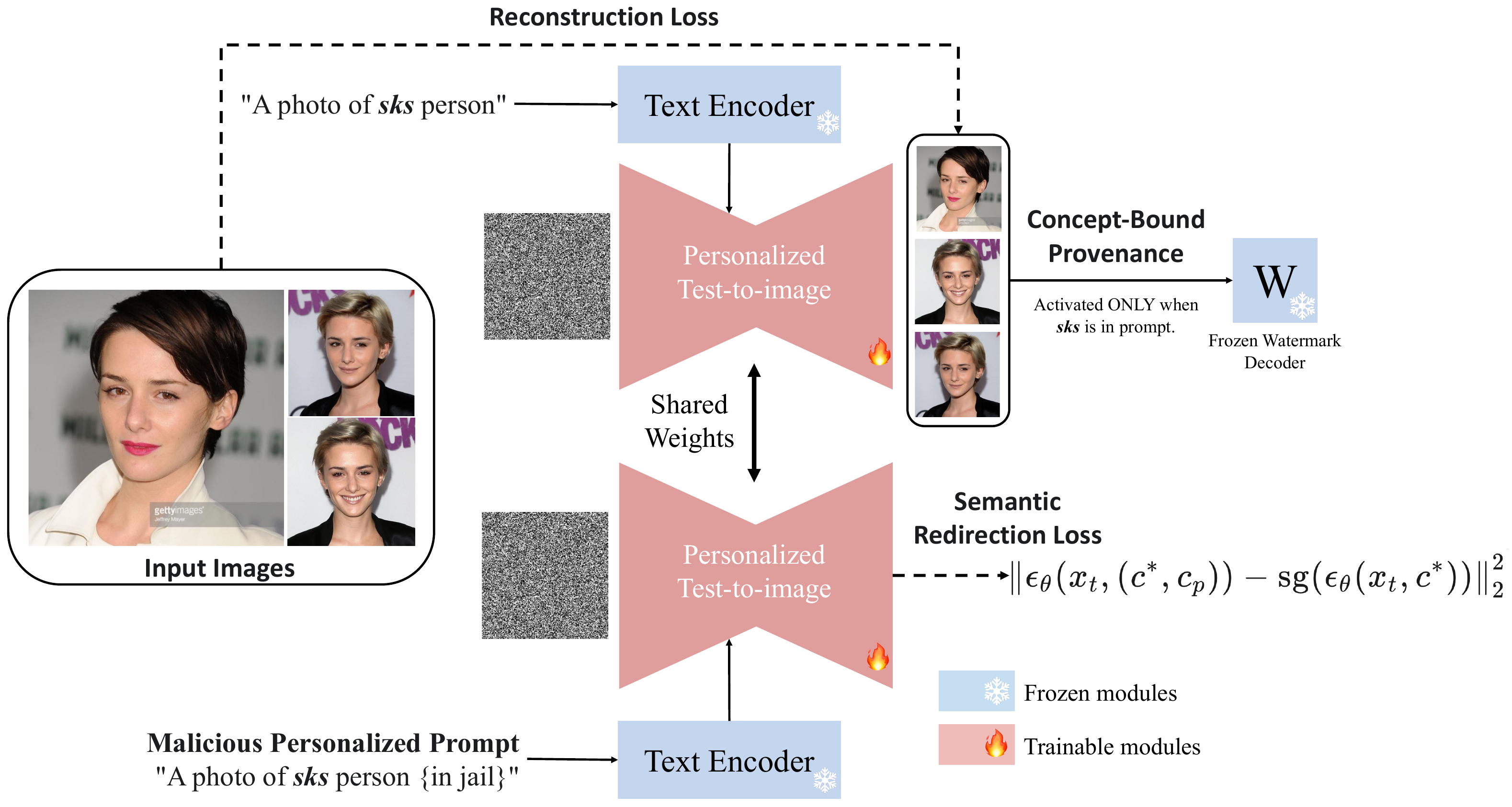}
    \caption{
        The IDENTITYGUARD fine-tuning framework. Our method trains a single Denoising U-Net using two conditional paths. (Top Path) For benign personalized prompts, our Concept-Bound Provenance is activated, embedding a watermark. (Bottom Path) For malicious prompts, our novel Semantic Redirection Loss is activated, redirecting the output towards a safe, identity-preserving result by aligning the noise predictions of the malicious and benign prompts.  Here, $c^*$ is the embedding for the personalized concept, $c_p$ is for the prohibited concept, and $\text{sg}(\cdot)$ is the stop-gradient operator.
    }
    \label{fig:method_overview}
\end{figure}

\texttt{IDENTITYGUARD} realizes our principle of Concept-Bound Security by integrating two conditional, context-aware mechanisms directly into the personalization fine-tuning loop, as illustrated in Figure~\ref{fig:method_overview}. The model learns from two parallel training paths acting on Denoising U-Net. The total objective augments the standard DreamBooth \cite{ruiz2023dreambooth} reconstruction loss, $\mathcal{L}_{\text{DB}}$, with our two novel security losses:
\begin{equation}
\mathcal{L}_{\text{total}} = \mathcal{L}_{\text{DB}} + \lambda_r \mathcal{L}_{\text{CIP}} + \lambda_w \mathcal{L}_{\text{WM}}
\end{equation}
where $\lambda_r$ and $\lambda_w$ are scalar weights. Unlike global penalties, our security losses activated intelligently based on the meaning of the prompt.

\subsection{Conditional Restriction via Semantic Redirection}
\label{ssec:restriction}

Generic erasure methods are a heavy-handed approach that causes collateral damage. We instead teach the model a more nuanced, conditional behavior: ``When you see personalized concept $c^*$ combined with a prohibited term, ignore the prohibited term and generate only $c^*$.''

We implement this behavior, which we call \textbf{Semantic Redirection}, through a novel training objective we term the \textbf{Conditional Identity-Preserving (CIP) Loss}. Let $c^*$ be the text embedding for the personalized concept and $c_p$ be the embedding for a prohibited concept. The CIP loss is activated only for malicious prompts and is formulated as:
\begin{equation}
\mathcal{L}_{\text{CIP}} = \mathbb{E}_{x_t, c_p} \left[ \left\| \epsilon_\theta(x_t, (c^*, c_p)) - \text{sg}(\epsilon_\theta(x_t, c^*)) \right\|_2^2 \right]
\end{equation}
where $x_t$ is the noised image, $\epsilon_\theta$ is the model's noise prediction, and $\text{sg}(\cdot)$ denotes the \texttt{stop\_gradient} operator.

The novelty of the CIP loss is defined by two key properties: it is both asymmetric and conditional. The asymmetric redirection—steering the prediction for the harmful combination towards the benign concept alone—is what faithfully preserves the user’s identity. Meanwhile, its conditional application is precisely what prevents the collateral damage that plagues context-blind methods.

\subsection{Concept-Bound Provenance}
\label{ssec:provenance}
To ensure meaningful provenance in personalized contexts, the signal must be robust and specific, serving as a distinctive signature that verifies the use of the personalized concept $c^*$.

To achieve this, we bind the watermark embedding process directly and exclusively to the personalized concept. As shown in Figure~\ref{fig:method_overview}, we use a pre-trained, \textbf{frozen} watermark decoder, $D_w$, to define a watermarking loss, $\mathcal{L}_{\text{WM}}$. This loss is computed only when the training prompt contains the personalized concept $c^*$:
\begin{equation}
\mathcal{L}_{\text{WM}} = 
\begin{cases} 
    \text{BCE}(D_w(x_{\text{gen}}), m) & \text{if } c^* \in \text{prompt} \\
    0 & \text{otherwise}
\end{cases}
\end{equation}
where $x_{\text{gen}}$ is the image and $m$ is the target $k$-bit message. 

This design represents a fundamental shift. The watermark is no longer a generic fingerprint on all model outputs. Instead, the personalized concept itself becomes the trigger for the watermark. This tight binding ensures both robustness and specificity, providing an unambiguous cryptographic link between an image and the identity used to create it.

\begin{figure*}[h]
    \centering
    \includegraphics[width=0.78\textwidth]{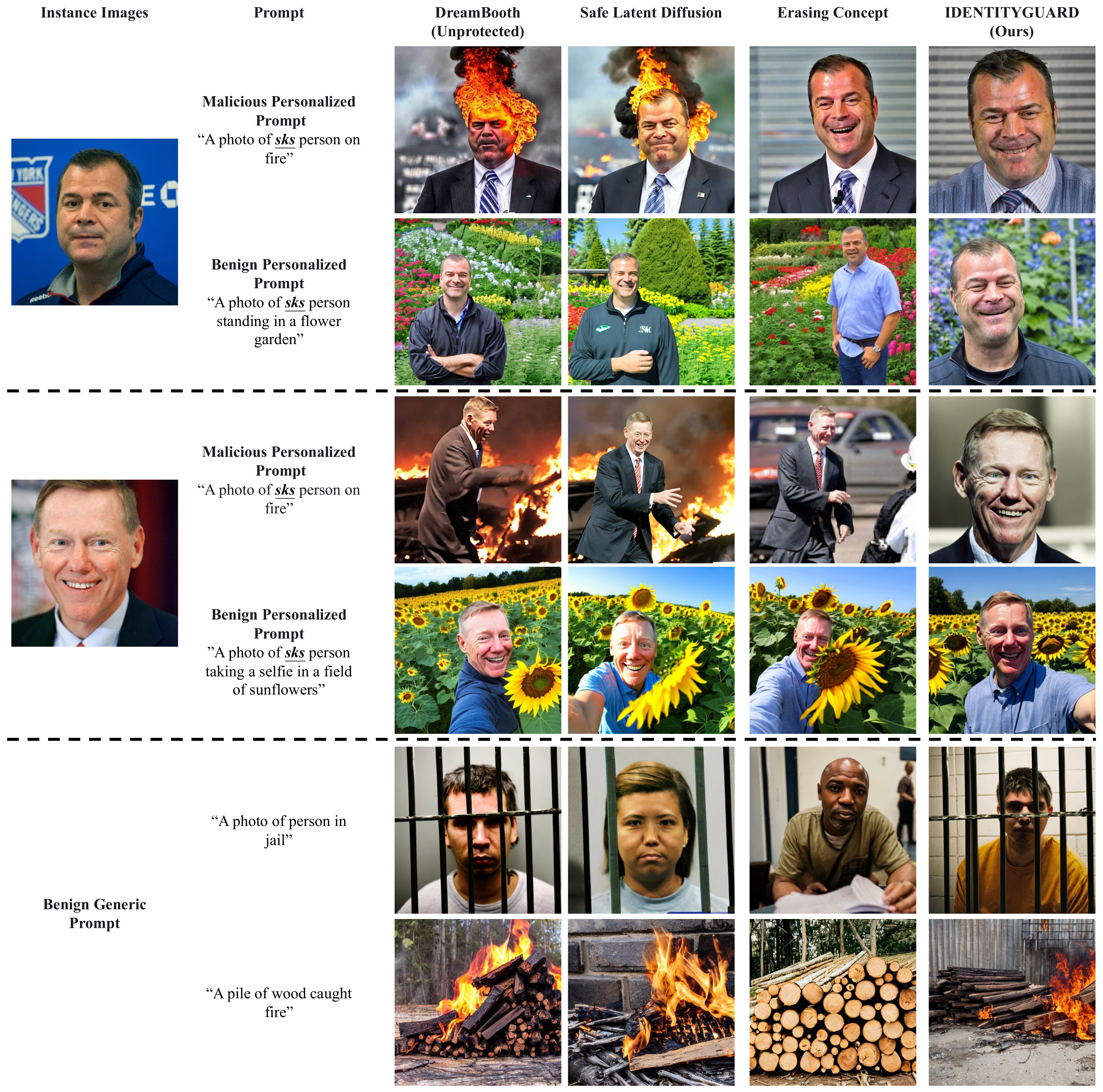}
    \caption{
         Qualitative analysis of our context-aware security. The key failure of generic methods like Erasing Concept(ESD) is revealed in the bottom rows: to provide protection in the personalized context (top rows), they are forced to inflict \textbf{catastrophic collateral damage}, globally erasing the concept of "fire" and failing to generate a simple campfire. In contrast, {IDENTITYGUARD}'s safeguard is intelligently bound {only} to the personalized identity, allowing it to preserve the concept for general use. 
    }
    \label{fig:qualitative_results}
\end{figure*}

\section{EXPERIMENTS}
\label{sec:experiments}
We conducted experiments to validate our main hypothesis that a context-aware Concept-Bound security paradigm is demonstrably superior to generic context-blind filters. We present a comprehensive comparison on core metrics, and then provide a deep dive into a real-world safety scenario.

\subsection{Experimental Setup}

\noindent\textbf{Implementation:} Our method is integrated into the DreamBooth fine-tuning process of a Stable Diffusion v2.1 model, with loss weights set to $\lambda_r = 0.2$ and $\lambda_w = 0.1$.

\noindent\textbf{Baselines:} We evaluate against methods representing the context-blind paradigm. For restriction, we use {Safe Latent Diffusion (SLD)}\cite{schramowski2023safe} and {Erasing Concepts (ESD)}\cite{gandikota2023erasing}. For provenance, we test {HiDDeN}\cite{zhu2018hidden}. 

\noindent\textbf{Metrics:} We measure image fidelity with {Fr\'echet Inception Distance (FID)} and {CLIP Score} on benign prompts. Restriction effectiveness is measured by {FID-Censored} and {CLIP-Censored} scores on malicious prompts. Watermark robustness is measured by {Bit Accuracy}. The term 'Censored' refers to prompts that contain prohibited semantics.

\subsection{Main Results: A Unified Analysis}

Our main results, consolidated in Table~\ref{tab:main_unified_results}, reveal the fundamental flaw of the generic security paradigm: it is forced to inflict collateral damage to provide security.

The unprotected DreamBooth model (FID 55.81) offers high fidelity but no protection. The generic safeguards attempt to fix this, but at a great cost. {Global Erasure (ESD)}, while providing strong restriction (CLIP-Censored 0.2093), suffers from precisely the collateral damage we hypothesized. As powerfully illustrated in our qualitative analysis (Figure~\ref{fig:qualitative_results}), ESD is forced to globally erase concepts, destroying the model's ability to generate benign images like a simple campfire. This forced trade-off makes it an impractical solution. Similarly, \textbf{post-hoc watermarking} is fundamentally incompatible with personalization, with a recovery rate barely better than random chance.

In stark contrast, our \textbf{Concept-Bound} framework, IDENTITYGUARD, is the only approach that avoids this dilemma. The table also serves as an ablation of our core mechanism, confirming that our final \texttt{Target} strategy is superior. Specifically, \texttt{Target} redirects the malicious prompt towards the benign personalized concept, whereas \texttt{Untarget} guides towards a null-text prompt, and \texttt{Conditioning} adds a preservation loss to ensure the blacklist concept remains usable when not combined with the personalized identity. Our final \texttt{Target} method achieves a state-of-the-art restriction score (CLIP-Censored \textbf{0.1919}) and near-perfect provenance (Bit Accuracy \textbf{97.1\%}), all while preserving the model's general utility and original fidelity.

\subsection{Case Study: Efficacy on a Critical Safety Threat}

To demonstrate the real-world impact of our method, we performed a case study on preventing the generation of nudity. As shown in Table~\ref{tab:nudity_case_study}, we used a dedicated nudity classifier (NudeNet)~\cite{bedapudi2019nudenet} to analyze 100 images generated from a malicious ``naked'' prompt.

The results are striking. The unprotected model produced 342 instances of exposed content. Even the strongest generic competitor, ESD, still had 46 detections, representing a significant safety failure. Our method, {IDENTITYGUARD}, provides a near-total solution, reducing the number of detections to just {2}. This order-of-magnitude improvement in a high-stakes scenario underscores the practical necessity of context-aware paradigms for building genuinely safe AI.

\noindent\textbf{Limitations and Future Work.}
Our experimental validation, while demonstrating the core principles of our paradigm, focuses on a curated set of concepts and baselines. We view this work as a strong proof-of-concept for context-aware security. Future work could explore scaling this approach to a broader range of generative architectures and more complex, open-ended blacklist definitions. 

\section{CONCLUSION}
\label{sec:conclusion}
In this work, we argued that the targeted risks of personalized generative models demand a targeted security response. We demonstrated that the prevailing paradigm of generic, context-blind filters is fundamentally flawed, forcing a false choice where security comes at the cost of unacceptable collateral damage to a model's broader utility.
As a solution, we proposed and validated a new paradigm: context-aware security, where safeguards are intrinsically bound to the concepts they protect. Our framework is a practical realization of this principle, successfully avoiding the collateral damage of older methods while providing both content restriction and robust, concept-specific traceability, demonstrates a more effective path toward building truly safe personalized AI.

\section{ACKNOWLEDGMENT}
This work was funded in part by the National Key R\&D Program of China under Grant 2022YFB3104300, the Jiangsu Provincial Natural Science Foundation of China under Grant BK20240291 and the Key Research and Development Programme of Ningbo’s “Science and Technology Innovation Yongjiang 2035” Plan under Grant 2025Z054.

\bibliographystyle{IEEEbib}
\bibliography{refs, refs2, strings}

\end{document}